\documentclass[twoside,12pt]{article}
\usepackage{epsfig}
\usepackage{graphicx}
\usepackage{amssymb}

\begin{document}
\title{Propagation of broad meson resonances in a BUU type transport model:
Application to di-electron production}
\author{{\sc H.W. Barz,$^1$ B. K\"ampfer,$^1$ Gy. Wolf,$^2$ M. Z\'et\'enyi$^2$}\\[3mm]
$^1$Forschungszentrum Rossendorf, Institut f\"ur Strahlenphysik,\\
PF 510119, 01314 Dresden, Germany\\[1mm]
$^2$KFKI RMKI, H-1525 Budapest, POB 49, Hungary} 
\maketitle
\begin{abstract}
We apply a BUU type transport model for the interpretation of
the di-electron invariant mass spectrum measured by the HADES collaboration for
the reaction \mbox{C(2 AGeV) + C}. Our model incorporates the propagation of broad
meson resonances emerging from the decay of baryon resonances.
\end{abstract}

\section{Introduction}

Di-electrons serve as direct probes of dense nuclear matter stages 
during the course of heavy-ion collisions. The superposition of various sources,
however, requires a deconvolution of the spectra by means of models.
Of essential interest are the contributions of the light vector mesons $\rho$ and $\omega$. 
The spectral functions of both mesons are expected to be modified in a strongly interacting
environment in accordance with chiral dynamics, QCD sum rules etc.\ \cite{RW}. 
After the first pioneering experiments with the DLS spectrometer \cite{DLS}
now improved measurements with HADES \cite{HADES} start to explore systematically 
the baryon-dense region accessible in fixed-target heavy-ion experiments at beam
energies in the few AGeV region. 

We apply here our BRoBUU transport model to the di-electron data measured by HADES in the
reaction C(2 AGeV) + C. Some features of the code are described in section 2, while
section 3 is devoted to the presentation of simulation results and the comparison to data.  
  
\section{BRoBUU code}

The BRoBUU computer code for heavy-ion collisions
developed by a Budapest-Rossendorf cooperation
solves the Boltzmann-\"Uhling-Uhlenbeck equation in the
quasi-particle \mbox{limit \cite{wolf93}} 
\begin{eqnarray}
\frac{\partial F}{\partial t} + \frac{\partial H}{\partial {\bf p}}
\frac{\partial F}{\partial {\bf x}} -  \frac{\partial H}{\partial {\bf x}}
\frac{\partial F}{\partial {\bf p}}  = {\cal C}, \quad 
H = \sqrt{(m_0+U({\bf p},{\bf x}))^2 + {\bf p}^2}
\end{eqnarray}
for the one-body distribution function $F({\bf x},{\bf p},t)$ of 
a certain hadron species.
This equation is applied to
the motion of different hadron species, each with rest mass $m_0$, 
in a momentum and density dependent mean field $U$.
The scalar mean field $U$ is chosen in such away that the Hamiltonian $H$
equals $H=\sqrt{m_0^2+{\bf p}^2}+U^{nr}$ with a usually in a non-relativistic
manner calculated potential $U^{nr}$
\begin{equation}
U^{nr} = A \frac{n}{n_0} + B \left( \frac{n}{n_0} \right)^\tau
+ C \frac{2}{n_0} \int \frac{d^3 p'}{(2\pi)^3} 
\frac{f_N(x,p')}{1 + \left( \frac{{\bf p} - {\bf p}'}{\Lambda} \right)^2},
\end{equation}
where the parameters $A$, $B$, $C$, $\tau$, $\Lambda$ define special types of 
potentials, while $n$,  $n_0$ and $f_N$ stand for the baryon number density, 
saturation density and nucleon distribution function.
The BRoBUU code propagates in the baryon sector the nucleons and
24 $\Delta$ and $N^*$ resonances and additionally 
$\pi,\eta,\sigma,\omega$ and $\rho$ mesons. 
Details will be reported elsewhere. 
Different particles species (each described by a corresponding distribution
$F$) are coupled by the collision integral ${\cal C}$ which 
also contains the \"Uhling-Uhlenbeck terms responsible
for Pauli blocking in the collision  as well as
particle creation and annihilation processes. 
The set of coupled Boltzmann-\"Uhling-Uhlenbeck equations is solved by using the
parallel-ensemble test-particle method, where each distribution function
is represented by 
$F = \sum\limits_{i=1}^{N_{test}}
\delta^{(3)} ({\bf x} - {\bf x}_i(t)) \delta^{(4)} (p - p_i(t))$.
This method transforms the 
partial differential--integro equations
into a set of ordinary differential equations (looking like equations of motion)
for a large number of test particles
simulating the ensemble averaging process for the respective function $F$.

Recently theoretical progress has been made in describing the in-medium properties
of particles starting from the Kadanoff-Baym equations \cite{Kadanoff} for
the Green functions of the particles. 
In the medium particles have a finite life time which is 
described  by  the width $\Gamma$ in the 
spectral function ${\cal A}$ of the particles.
The spectral function being the imaginary part of the retarded propagator
in the Kadanoff-Baym equations, $-2 Im G^{ret}(x,p)$, is essentially defined
by the self-energies $\Sigma$ of the particle in the medium. For bosons it reads
\begin{equation}
{\cal A }(p) = \frac{\hat{\Gamma}}{(E^2 -{\bf p}^2-m_0^2-          
{\rm Re} \Sigma^{ret})^2 + \frac14 \hat{\Gamma}^2},
\end{equation}
where Im$\Sigma^{ret} \equiv \frac12 \hat \Gamma$ and Re$\Sigma^{ret}$ 
depend on $E, {\bf p}$ and the local medium properties.
The spectral function can significantly 
change during the heavy-ion collision process and can be simulated by an
ensemble of test particles with different masses. 
The change of the spectral function
is now given by time variation of the test particle mass 
$m_i$ \cite{Cassing-Juchem00,Leupold00}. For bosons this additional equation reads
\begin{eqnarray} 
\frac{dm_i^2}{dt} \,\approx\, (\frac{\delta}{\delta t} {Re \Sigma^{ret}} +
\frac{m_i^2 - m_0^2 - Re \Sigma^{ret}}{\hat \Gamma}{\frac{\delta}{\delta t} {\hat \Gamma}}),
\label{eq.2}
\end{eqnarray}  
where the values of the selfenergy $\Sigma^{ret}$ are taken at the positions of
the test particles $i$, and $\delta/\delta t$  stands for the comoving time 
derivation. The real part of  the selfenergy is related to the 
mean field $U$. Eq.~(\ref{eq.2}) is actually a corollary of three equations
describing the propagation of energy, three-momentum and position 
of a test particle. 
This equation ensures that resonances
are propagated towards their vacuum spectral functions at freeze-out 
\cite{Bratkovskaya}.
This technique, allowing for a consistent 
propagation of broad resonances,
is applied in the BRoBUU code for calculating the di-electron 
emission of $\omega$ and $\rho$ mesons. 

Besides the propagation
of broad resonances the spectral function also controls the 
production of mesons.
Vector mesons ($V$) are essentially created by the decays 
of baryon resonances ($R$).  
The mass distribution reads
\begin{equation}
\frac{dN^{R \to N V}}{dm_V} \propto {\cal A}(m_V) \,m_V \, (m_R^2+m_N^2-m_V^2) \,
\lambda^{1/2}(m_R^2,m_N^2,m_V^2),
\end{equation}
where $\lambda(a^2,b^2,c^2) = (a^2+b^2-c^2)^2 -4a^2b^2$ denotes the triangular
function. The resonances $R$ is created in nucleon-nucleon collisions $NN \to NR$
and meson-nucleon collisions, $M N \to R$. $R$ may more generically decay as
$R \to R' M$ into other channels ($R'$=$\Delta(1232)$, $N(1440)$, $N(1520)$,
$N(1535)$).
Resonance parameters for $NN\leftrightarrow NR$ and $MN\leftrightarrow R$
are fitted to available data for meson production in nucleon-nucleon 
and pion-nucleon reactions \cite{wolf97}. 

In our calculations we  employ a simple form of the selfenergy:
\begin{eqnarray} \label{areal}
{\rm Re} \Sigma^{ret}_V & = & 2 m_V \Delta m_V \frac{n}{n_0},\\
\label{aimag}
{\rm Im} \Sigma^{ret}_V & = & m_V (\Gamma^{vac}_V + 
\frac{n v \sigma_V}{\sqrt{1-v^2}}).
\end{eqnarray}
Eq.~(\ref{areal}) describes schematically a mass shift proportional to the
density $n$ of the surrounding matter related to the normal matter density $n_0$. 
The effect of the collision broadening is given in Eq.~(\ref{aimag}) which 
depends on density, relative velocity $v$ and the cross section $\sigma_V$
of the vector meson in matter. This cross section 
is calculated via the Breit-Wigner formula
\begin{equation}
\sigma_V =  \frac{4\pi}{q_{in}^2} \sum\limits_R \frac{2J_R+1}{3(2J_i+1)} 
\frac{s \Gamma_{V,R}\Gamma^{tot}_R}{(s-m_R^2)^2 +              
s(\Gamma^{tot}_R)^2}
\end{equation}
for forming resonances with masses $m_R$, angular momenta $J_{R}$, partial widths
$\Gamma_{V,R}$, total widths $\Gamma^{tot}_R$ with energy $\sqrt{s}$ and relative
momentum $q_{in}$ in the entrance channel.
In vacuum the baryon density $n$ vanishes and the resulting spectral
function${\cal A}_{vac}$ is solely determined 
by the energy dependent width $\Gamma_V^{vac}$.

The di-electron production from vector meson decays $V \to e^+ e^-$
is calculated by integrating the local decay probabilities 
along their paths in the collision.  
The subleading so-called direct channel
$\pi \pi \to \rho \to e^+ e^-$ is treated with the cross section
\begin{equation}
\sigma (M) = \sigma_{vac} \frac{m_\rho}{m_{\rho, 0}}
\frac{{\cal A}}{{\cal A}_{vac}}, \quad
\sigma_{vac} = \frac{4\pi}{3} \left( \frac{\alpha}{M}\right)^2
\sqrt{1 - \frac{4 m_\pi^2}{M^2}} \,
\frac{\tilde m_\rho^4}{(M^2 - \hat m_\rho)^2 -
(\tilde m_\rho \tilde\Gamma)^2}
\end{equation}
with $m_\rho = m_{\rho,0} + \Delta m_\rho n /n_0$, 
$\tilde m_\rho = 775$ MeV,
$\hat m_\rho = 761$ MeV,
$\tilde \Gamma_\rho = 118$ MeV. 
Analog considerations apply to the channel
$\pi \rho \to \omega$.

We also include into our simulations a bremsstrahlung contribution which is guided
by a one-boson exchange model adjusted to $pp$ virtual bremsstrahlung 
and transferred
to $pn$ virtual bremsstrahlung \cite{Kaptari}. Actually we use
$\frac{d \sigma_{pn}}{d M} = \frac{4 \pi}{M} \, \sigma_\perp \, 
\frac{\alpha^2}{6 \pi^3} \int dq q^2/q_0^3  R_2(\bar s)/ R_2(s)$ 
and $\sigma_{pp} = 0$ with 
$\sigma_\perp = \sigma_{pn,tot} (s) \frac{s - 4 m_N^2}{2 m_N^2}$,
where $M$ is again the $e^+e^-$ invariant mass, $R_2$ denotes 
the two-particle phase space volume, $\sqrt{s}$ stands for the c.m.s.\
energy in a nucleon-nucleon collision, $\bar s$ is the reduced energy squared
after the dilepton emission, and $\sigma_{pn,tot} (s)$ is the
corresponding total cross section. 

\section{Results and comparison to data}

We employ the above described code for the reaction C(2 AGeV) + C, where first
data from HADES \cite{Sturm} are at our disposal. 
In the present explorative study
we are going to contrast simulations with and without medium modifications
of $\rho$ and $\omega$ mesons to elucidate to which degree medium effects
can become visible in the light collision system under consideration.
In doing so we use fairly schematic medium effects condensed in a ''mass shift''
of $\Delta m_\omega = - 50$ MeV for the $\omega$ meson. 
Such a shift is suggested by recent CB-TAPS
data \cite{Trnka}. The use of QCD sum rules \cite{QCDSR} then can be used to
translate this shift into a significantly larger shift for the $\rho$ 
meson (dictated essentially by the Landau damping term); 
we use here $\Delta m_\rho = - 200$ MeV. 

\begin{figure}[!htb]
\center
\epsfig{file=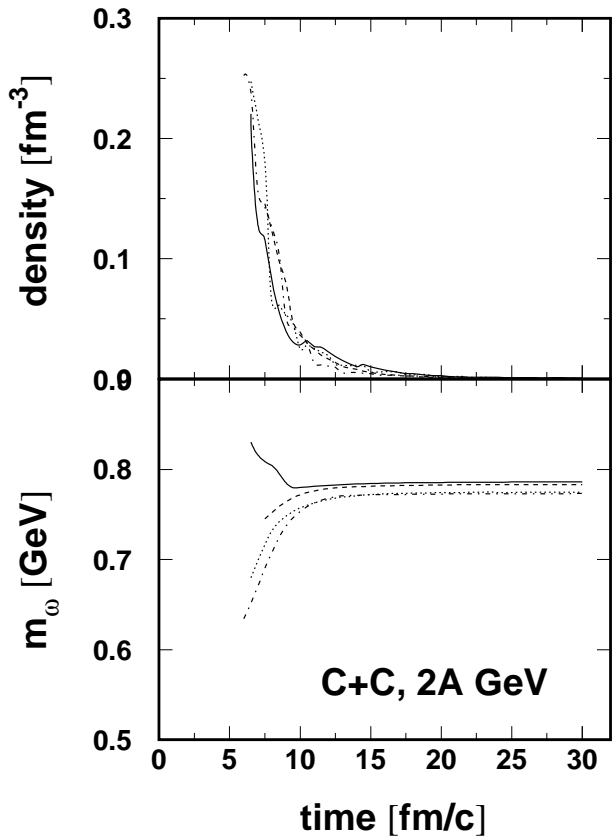,width=0.5\linewidth,angle=0}
\caption{\it Time evolution of the masses (lower panel) and the surrounding
baryon densities (upper panel) of four randomly selected test particles.} 
\label{fig_evol}
\end{figure}

\begin{figure}[!htb]
\center
\epsfig{file=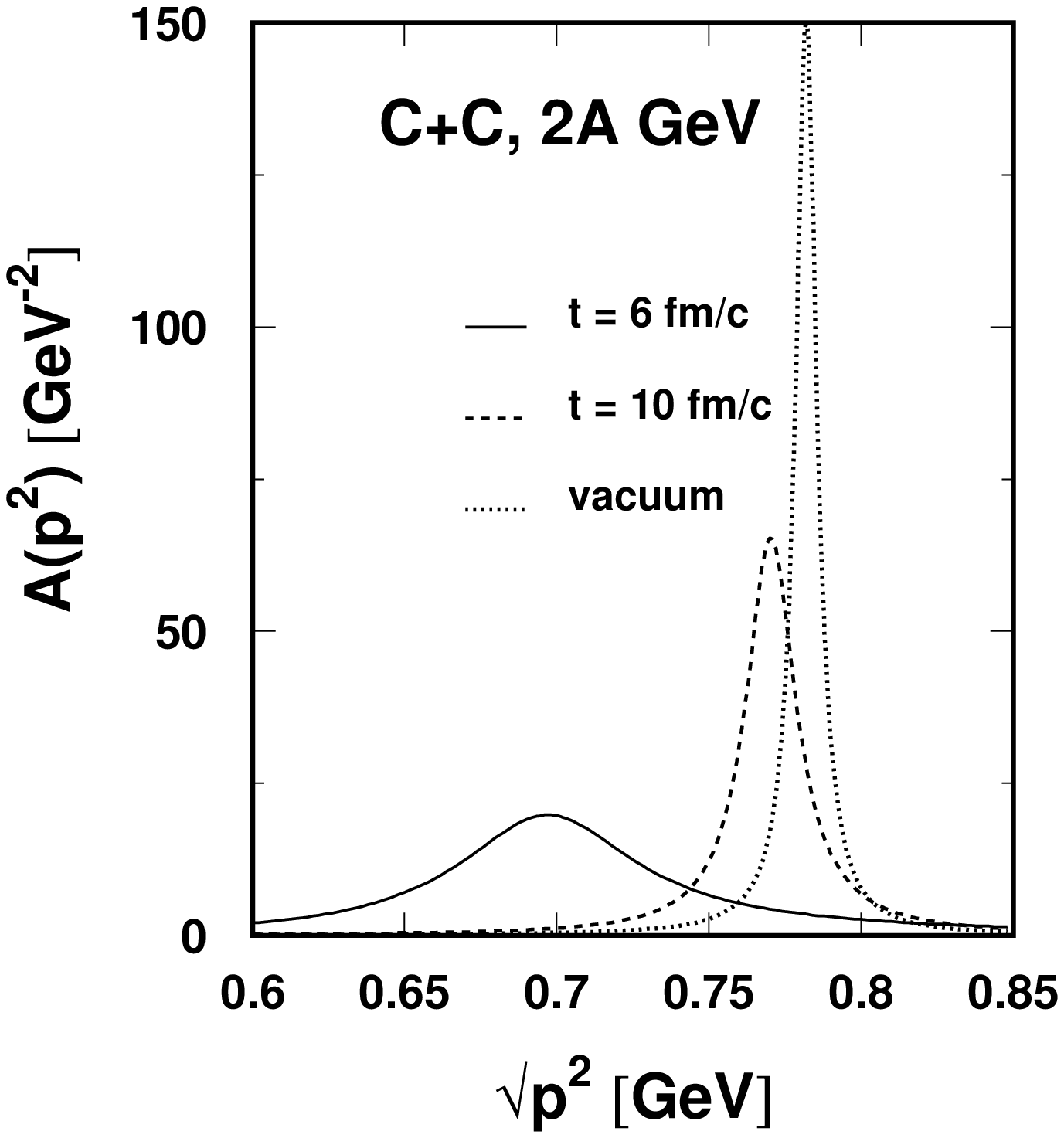,width=0.5\linewidth,angle=0}
\caption{\it Spectral functions for the $\omega$ meson at densities
felt by the test particle displayed by the dash-dotted line 
in Fig.~\ref{fig_evol} at collision times of 6 fm/c, 10 fm/c
and in the vacuum, respectively.} 
\label{fig_spec}
\end{figure}

To illustrate the time evolution of the test particles we have shown in
the lower part of Fig.~\ref{fig_evol} the masses of four test particles
randomly selected 
out of the ensemble representing the spectral function of an $\omega$
meson as a function of the collision time. (These curves were obtained
by artificially switching off the decay of the $\omega$ mesons. Otherwise
they would decay in a very short time in the high density region. Thus,
particles with masses strongly deviating from their peak mass have 
little chance to radiate di-electrons off.)   
The upper part shows the respective local densities of the matter 
surrounding these particles. 
The tendency to acquire their vacuum mass starting from very different
initial masses is clearly seen. For the test particle depicted by the
dash-dotted curve which moves with a velocity of 0.6 c  we have also 
shown the spectral function of the $\omega$ meson
at the collision times of 6 fm/c and 10 fm/c together with the vacuum function
in Fig.~\ref{fig_spec}. 

\begin{figure}[!htb]
\center
\epsfig{file=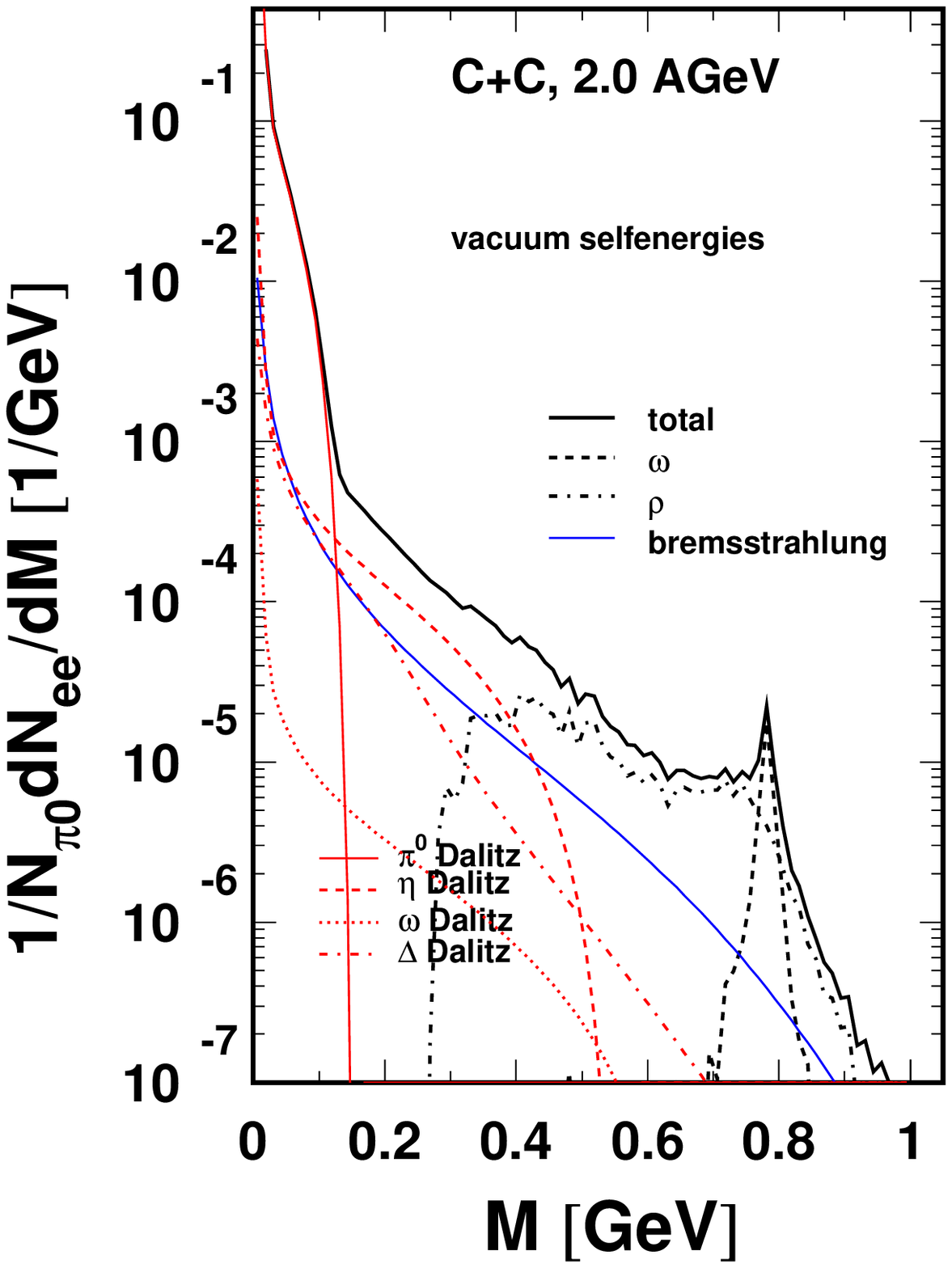,width=0.42\linewidth,angle=0} \hfill
\epsfig{file=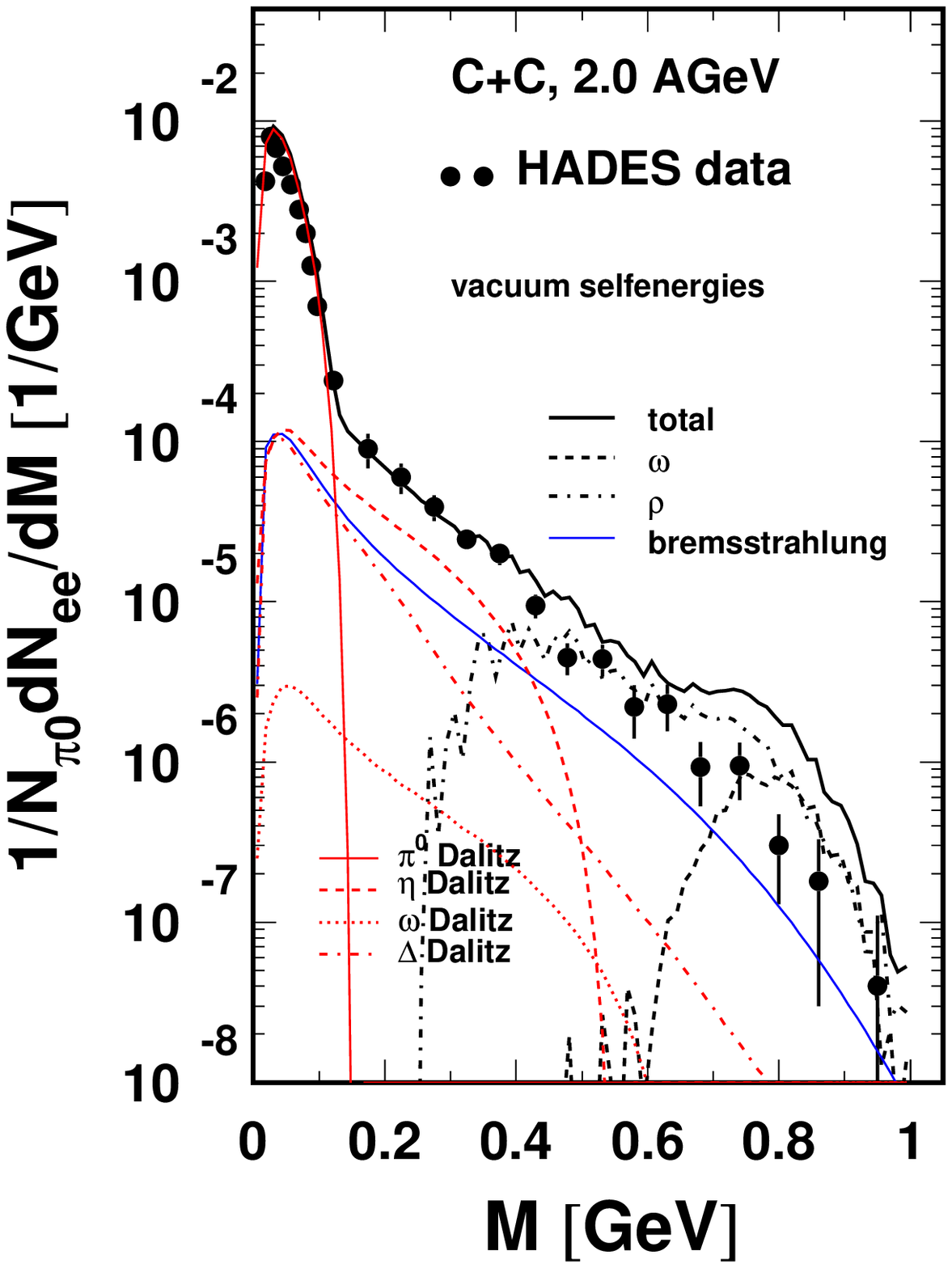,width=0.42\linewidth,angle=0}
\caption{\it Various sources of the di-electron invariant mass spectrum
for C(2 AGeV) + C. Left panel: without filter. Right panel: with experimental filter
and compared to HADES data \cite{Sturm}. Vacuum selfenergies are used.
\label{fig_HADES_vac}}
\vskip 9mm
\center
\epsfig{file=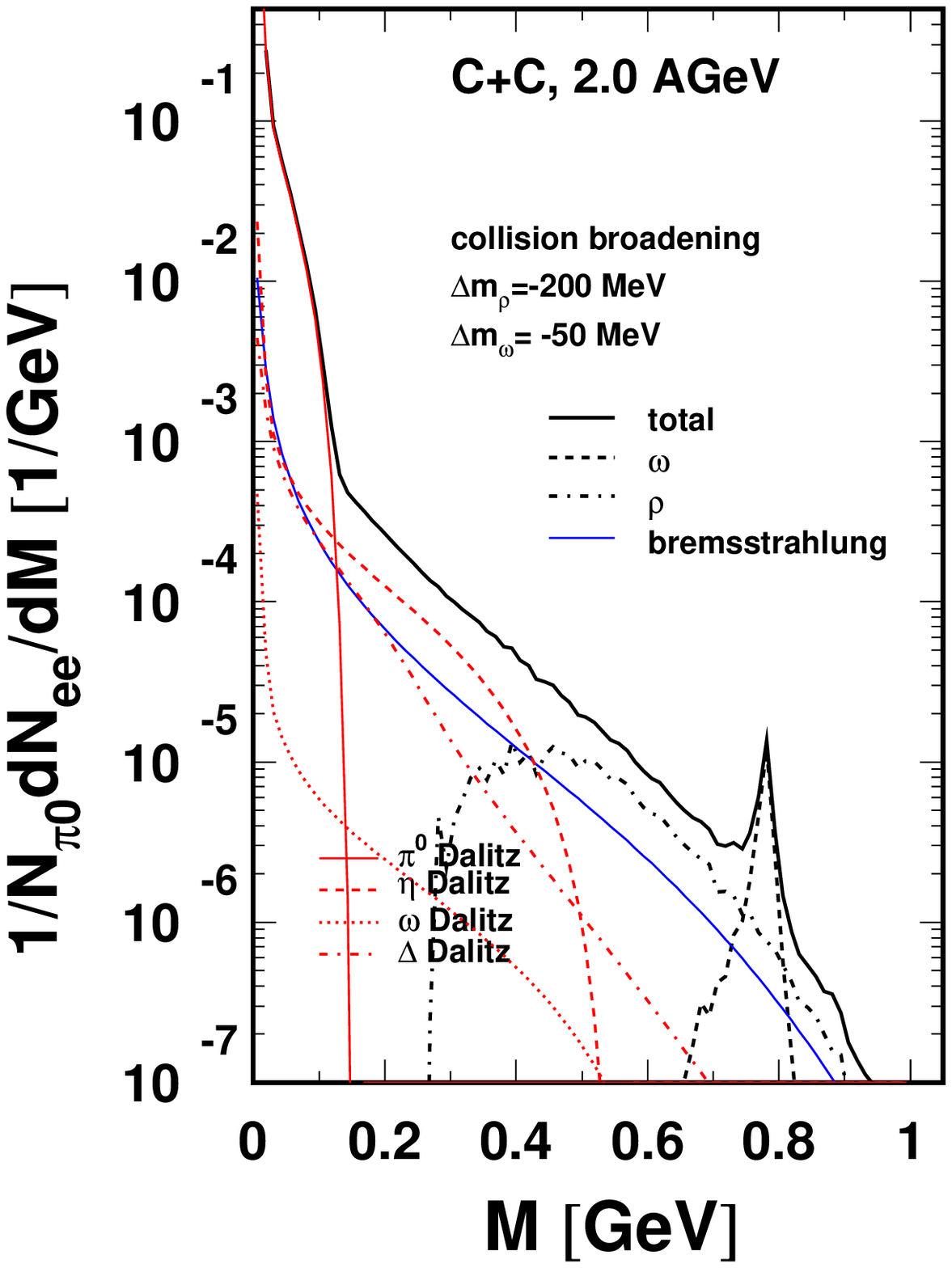,width=0.42\linewidth,angle=0} \hfill
\epsfig{file=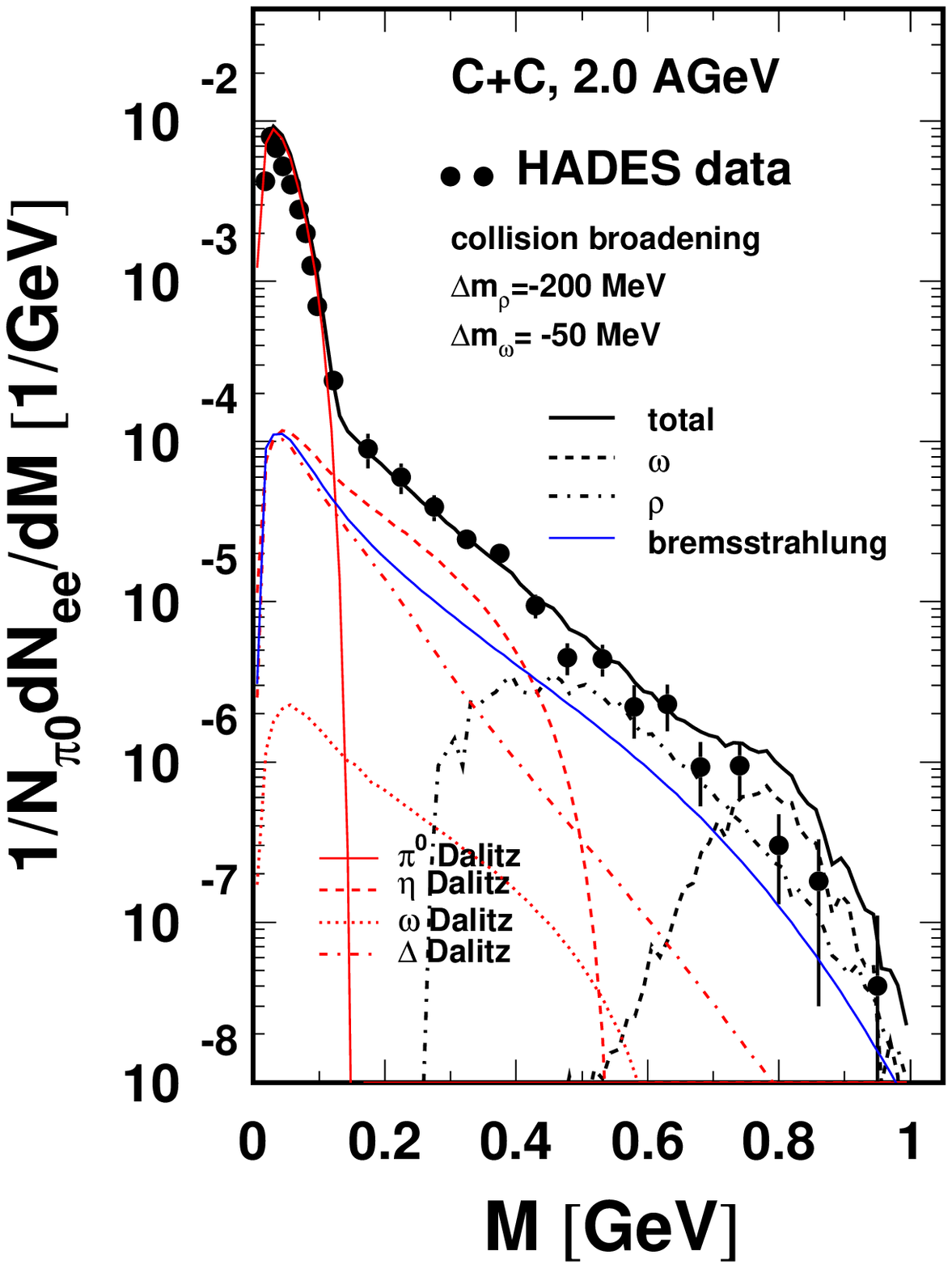,width=0.42\linewidth,angle=0}
\caption{\it The same as Fig.~\ref{fig_HADES_vac} but with modified 
selfenergies of $\rho$ and $\omega$ mesons
as described in text.
\label{fig_HADES_med}}
\end{figure}

The obtained di-electron spectra are represented in Figs.~\ref{fig_HADES_vac} and 
\ref{fig_HADES_med}.
In Fig.~\ref{fig_HADES_vac} the vacuum parameters are employed
while Fig.~\ref{fig_HADES_med} shows the results obtained by including the above 
described medium modifications of $\rho$ and $\omega$ mesons.
For comparison with the data the HADES filter has been applied
accounting for the geometrical acceptance,
momentum cuts and pair kinematics. The filter causes a reduction of the strength
and a smearing of the invariant masses of the di-electrons.
The result of this filtering is always shown on the right hand panel of the figures.

In these figures we show different contribution to the 
di-electron rate. Important
di-electron sources are $\pi^0$ and $\eta$ Dalitz decays 
which are 
proportional to the multiplicities of their parents.
The TAPS collaboration has measured \cite{TAPS} the $\pi^0$ and $\eta$ production 
cross sections of 707$\pm$72 mb and 25$\pm$4 mb for the same system and same bean energy.
These values have to be compared
to our calculations yielding 700 mb and 20 mb. While the value for pion production 
is in good agreement, there is a slight underestimation of the $\eta$ production.
(Note that the presently employed cross sections rely on a global fit of many
elementary reactions which is not optimized for the $\eta$ channel.) 
The Dalitz decays of $\rho$ and $\omega$ mesons and nucleon resonances 
do not  contribute essentially.
The employed elementary cross sections for $pp \to pp \omega$ and $pn \to pn \omega$
agree with the ones in \cite{Kaptari_omega} which are adjusted to near-threshold
data for $pp \to pp \omega$.  
One recognizes comparing Fig.~\ref{fig_HADES_vac}
and \ref{fig_HADES_med} that the medium modification due
to collision broadening diminishes the yield of  $\rho$ 
di-electrons. Still an indication of a shoulder is expected from
the calculations. Our calculations  do not show a considerable effect 
of a $\rho$ mass shift, however this finding depends 
sensitively on our elementary $\rho$ production rates employed in the
calculations. Since the fine structure is not yet resolved in the data 
a conclusive decision cannot be made at present.

\section{Summary}

In summary we have compared results of a transport code to recent measurements
of the di-electron invariant mass spectrum for the reaction C(2 AGeV) + C.
The code incorporates the production and propagation of broad resonances,
in particular $\rho$ and $\omega$ mesons. Including even fairly schematically
parameterized in-medium modifications of these vector mesons the emerging
differences to results which employ vacuum parameters show up in a better
agreement with the data.  

{\it Acknowledgements:} We gratefully acknowledge the continuous information by the
HADES collaboration, in particular R.\ Holzmann for delivering and assisting
us in using the acceptance filter routines HAFT. The work is supported by the German 
\mbox{BMBF 06DR121} and the Hungarian OTKA T46347 and T48833.


\end{document}